\documentclass[twocolumn,english,aps,floatfix,superscriptaddress,showpacs,amsfonts,amssymb,superscriptaddress]{revtex4}
\usepackage{color}

\usepackage{graphicx}
\usepackage{amsmath}
\usepackage{latexsym}
\usepackage{epstopdf}
\usepackage{amsmath}
\usepackage{amssymb,amsmath}
\usepackage{multirow}
\usepackage{mathtools}
\newcommand{\beq}{\begin{equation}}
\newcommand{\eeq}{\end{equation}}

\def\ket#1{|#1\rangle}

\newcommand{\eq}{\begin{equation}}
\newcommand{\en}{\end{equation}}
\newcommand{\ear}{\begin{eqnarray}}
\newcommand{\rae}{\end{eqnarray}}

\newcommand{\tr}{{\rm tr}\,}
\newcommand{\Tr}{{\rm tr}^{~}}

\def\ket#1{|#1\rangle}

\begin{document}
\title{Post measurement bipartite entanglement entropy  in conformal field theories}

\author{M.~A.~Rajabpour}
\affiliation{ Instituto de F\'isica, Universidade Federal Fluminense, Av. Gal. Milton Tavares de Souza s/n, Gragoat\'a, 24210-346, Niter\'oi, RJ, Brazil}

\date{\today{}}

\begin{abstract}
We derive exact formulas for bipartite von Neumann entanglement entropy after partial projective local measurement in $1+1$ 
dimensional conformal field theories with periodic and open boundary conditions. After
defining the set up we will check numerically the validity of our results in the case of Klein-Gordon field theory (coupled harmonic
oscillators) and spin-$1/2$ XX chain in a magnetic field. The agreement between analytical results and the numerical calculations is very good. 
We also find a lower bound for localizable entanglement in coupled harmonic oscillators.

\end{abstract}
\maketitle

In the last couple of decades bipartite entanglement entropy
attracted a lot of attention in the high energy physics and the many body condensed matter physics mostly due to
the area law property of bipartite von Neumann entanglement entropy which is the most celebrated measure of entanglement \cite{EE,Srednicki}.
In one dimensional quantum systems while the area law is usually valid just for massive (gapped) systems 
for the critical systems the entanglement entropy of subsystem with the size $l$ of the ground state follows the logarithmic formula
$S=\frac{c}{6}\log l$, where $c$ is the central charge of the underlying conformal field theory (CFT) describing the 
low energy behavior of the critical system \cite{Holzey1994,Viadal, Calabrese2004}. Since the central charge of the system
usually can fix the universality class of the system calculating entanglement entropy for a system
can give a lot of insight about the possible universality class of the system. In particular, since numerical calculation
of the entanglement entropy by using the techniques of DMRG in one dimension is now a well-known method \cite{Schol} one can easily
study the critical and non-critical properties of the quantum systems in one dimension by studying the entanglement entropy.
It is worth mentioning that entanglement entropy is not the only quantity which gives directly 
the central charge of the system. Among other measures one can name, mutual informations \cite{AR2013,Stephan2013, AR2015} and 
quantum contours \cite{Vidal2014}, see also other related works \cite{CCT,Melko2014}. The former one is based on local measurements in particular basis, 
so called conformal basis \cite{AR2013,AR2015}.
One of the simple properties of the entanglement entropy which makes it more appealing in numerical calculations with respect to other measures
such as Shannon and R\'enyi mutual informations is that it is completely independent of the basis that we write the wave function. However,
since it is a very non-local quantity it is not an easy quantity to be measured in the experiment, for recent developments see 
\cite{Cardy2011,Demler2012,Zoller2012,Greiner2015}. 
Although bipartite entanglement entropy has been studied in length in many quantum systems there are few studies regarding entanglement in multipartite 
systems \cite{Cirac2012}. One of a few entanglement measures regarding tripartite systems is localizable entanglement introduced in \cite{Verstraete2004},
see \cite{EE} for review. The localizable entanglement is defined as the maximal amount of entanglement that
can be localized, on average, by doing local measurements in part of the system. In other words the localizable entanglement
between the two parts $B$ and $\bar{B}$ after doing local measurement in the rest of the system $A$ is defined as
\begin{eqnarray}\label{localizable}
E_{loc}(B,\bar{B})=sup_{\mathcal{E}}\sum_K p_K E(\ket{\psi_{K}}_{B\bar{B}})
\end{eqnarray}
where $\mathcal{E}$ is the set of all possible outcomes ($(p_K,E(\ket{\psi_{K}}_{B\bar{B}}$) of the measurements and $E$ is the chosen entanglement measure.
Because of the maximization over all the possible local measurements the localizable entanglement is a very difficult quantity to calculate.
However, in those  cases that $B$ and $\bar{B}$ are single spins and $E(.)$ is the concurrence one can derive interesting lower bounds to the localizable
entanglement, see \cite{EE} and references therein and for other related works see also \cite{localizable E references}. The quantity has been also
measured experimentally in a system of two photons in a noisy surrounding in \cite{localizable Experiment}.
 Due to the complexity of the definition of localizable entanglement,
 to the best of our knowledge, the case of $B$ and $\bar{B}$  not being just single particles but many particle subsystems has not been addressed in the literature.
 In this work instead of calculating the very complicated quantity of localizable entanglement we calculate 
 $E(\ket{\psi_{K}}_{B\bar{B}})$ appearing in the definition (\ref{localizable}) for particular natural basis. In the case of harmonic oscillators
 our exact results are useful to find lower bound for the localizable entanglement.

 Consider the ground state of 
 a many body system (for example a spin chain) written in particular basis, i.e. $\ket{\psi_g}=\sum_{I}a_I|I>$, where for example
 in the case of spin chain $I$ are all the possible spin configurations in the $\sigma^z$ direction.
 If we do local projective
 measurement  of a local quantity (for example $\sigma^z$ in the spin chain) in some part of the system that part will take definite
 spin direction and will be disentangled 
 from the rest. In other words one can write $\ket{\psi'_g}=\sum_{I}a_IM_K\ket{I}=\sum_{J}a_{JK}\ket{J}\ket{K}$, where $M_K$ is the projective
 measurement operator of the subsystem with the outcome $K$ and the sum is now over the spins of the rest of the system. Notice that if we start with the ground state of a spin chain the
 resulting state after measurement
 is nothing to do with the ground state of the rest of the system.
 The  entanglement content of the new wave function will be dependent on the basis that we chose for our measurement and of 
 course
 to the outcome of the measurement, in other words, the entanglement content of the $\ket{\psi'_g}$ is dependent on $M_K$. 
 In a recent development \cite{AR2013} it is shown that although, in general, doing measurement in 
 arbitrary basis is not compatible with CFT set up there are some natural basis (conformal basis) that are closely related to boundary CFT. 
  In other words
 if one makes a measurement in these basis one can still hope to preserve some of the universal
 properties of the system and be able to calculate the entanglement entropy.  We noticed that
 engineering quantum spin chains and making projective measurement on natural local basis is now possible with optical lattices and  ion trapping techniques, see \cite{experiments} 
 and references therein. However, calculating entanglement entropy after partial local measurement in many 
 body quantum chains has not been investigated yet.
 Analytical calculation of bipartite
 von Neumann entanglement entropy after projective measurement in natural conformal basis is the main purpose of this article. To do that we first
 define our set up and fix our assumptions and then we will analytically calculate  entanglement entropy using twist operators of CFT.
Exact formulas will be derived and then we will check the validity of our results by explicit examples in the field theories such as
Klein-Gordon field theory (coupled
harmonic oscillators) and in quantum spin chains such as, XX model in a magnetic field.

\section{Bipartite entanglement entropy after partial measurement in CFT.}

The basic setup for our problem is as follows: consider the ground state of a periodic critical system in one dimension with the total size $L$.
Then we make a projective local measurement in the subsystem $A$ with the length $s$ of this system in particular local basis. After
the measurement the subsystem $A$ will be decoupled (unentangled) from the complement $\bar{A}$. However, the wave function of the subsystem $\bar{A}$
after measurement, i.e. $\psi_{\bar{A}}$, is highly entangled. In other words if we take a subsystem $B\subset \bar{A}$ it will be entangled with $\bar{B}$, where
$B\cup\bar{B}=\bar{A}$. From now on we will take the length of the subsystem $B$ equal to $l$ and for simplicity we will take $B$ and  
$\bar{B}$ in a way that they are adjacent simply connected domains. Here we are interested to calculate the entanglement
entropy of the subsystem $B$ with respect to $\bar{B}$, i.e. $S_B$, in a critical quantum chain. To calculate such a quantity it is first useful
to think about the Euclidean version of the system after measurement. For a periodic boundary conditions
\begin{figure} [htb] \label{fig1}
\centering
\includegraphics[width=0.3\textwidth]{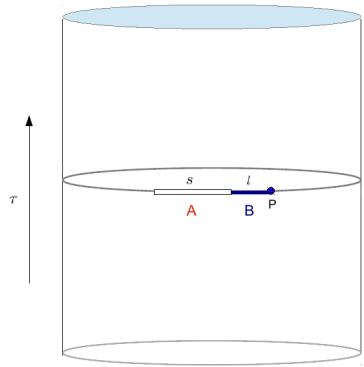}
\caption{Euclidean version of the quantum  chain with total length $L$. The removed slit domain $A$ has length $s$ and we are interested in calculating
 entanglement entropy of the region $B$ with length $s$ with the complement in the quantum spin chain. The twist operator can be put at point $P$.} 
\end{figure}using the transfer matrix approach one can simply map the system after measurement to a cylinder with a slit on it as it is demonstrated in Figure ~1, 
 see \cite{Stephan2013}. In general this system is not necessarily an example of boundary CFT except for those cases that the subsystem $A$ after measurement
picks up a configuration which renormalizes to a boundary CFT. We will discuss this important issue later in the case of critical spin chains. 
Now if we consider
that the outcome of the measurement leads us to a boundary CFT, then based on the Cardy-Calabrese argument \cite{Calabrese2004} the calculation of 
the entanglement entropy
$S_B$ should boil down to the calculation of the one point correlation function of twist operator sitting on the border between $B$ and $\bar{B}$ on the cylinder.
In other words we have:
\begin{eqnarray}\label{EE twist}
S_B=-\lim_{n\to 1}\frac{\partial}{\partial n}\Tr \rho_B^n=-\lim_{n\to 1}\frac{\partial}{\partial n}<\mathcal{T}_n(P)>_{_{cyl/slit}},
\end{eqnarray}
where $\mathcal{T}_n$ is the twist operator with conformal weight $h_n=\frac{cn}{24}(1-\frac{1}{n^2})$ and $c$ is the central
charge. The one point function
of the twist operator on the cylinder with a slit can be easily found by mapping the system to the upper half plane by using the map
$z(w)=\sqrt{\frac{\sin \frac{\pi}{2L}(s+2w)}{\sin \frac{\pi}{2L}(s-2w)}}$ and the following well-known formula of CFT \cite{BPZ}:
\begin{eqnarray}\label{conformal mapping}
<\mathcal{T}_n>_{_{cyl/slit}}=(\partial_wz)^{2h}<\mathcal{T}_n>_{\mathcal{H}},
\end{eqnarray}
where $<\mathcal{T}_n>_{\mathcal{H}}=\Big{(}\frac{a}{2z(s/2+l))}\Big{)}^{2h_n}$ is the one point function on the upper half plane $\mathcal{H}$. $a$ is 
the lattice spacing and we took the coordinate of the twist operator at $\frac{s}{2}+l$.  Putting all 
together we will have
\begin{eqnarray}\label{SB for PBC}
S_B=\frac{c}{6}\ln \Big{(}\frac{L}{\pi}\frac{\sin\frac{\pi}{L}(l+s)\sin\frac{\pi}{L}l}{a\sin\frac{\pi}{L}s}\Big{)}+\gamma_1+...,
\end{eqnarray}

where $\gamma_1$ is a constant and the dots are subleading terms. Notice that in the above formula $s$ can not go to zero because cylinder with a slit is topologically
different from cylinder. To get the result for before measurement case one can simply put $s=a$ which is the smallest scale
in the system, then we will have $S_B=\frac{c}{3}\ln(\frac{L}{a\pi}\sin\frac{\pi l}{L})$ which is the well-known result for the bipartite entanglement entropy in
CFT \cite{Calabrese2004}. In the next sections we will verify the validity of the above equation in two important cases:
Klein-Gordon field theory (harmonic oscillators)
and the critical XX model.

\section{Harmonic oscillators.}

In this section we would like to calculate bipartite entanglement entropy of the ground state
of coupled harmonic oscillators after measuring the position of a string of oscillators.
Consider the Hamiltonian of $L$-coupled harmonic 
oscillators, with coordinates $\phi_1,\ldots,\phi_L$ and conjugated momenta 
$\pi_1,\ldots,\pi_L$:

\begin{equation}\label{harmonicOsc}
\mathcal{H}=\frac{1}{2}\sum_{n=1}^{L}\pi_n^2+\frac{1}{2} \sum_{n,n^\prime=1}^{L}\phi_{n} K_{nn^\prime}\phi_{n^\prime}~.
\end{equation}
The ground state of the above Hamiltonian has the following form
\begin{equation}\label{GroundSwave}
\Psi_0=(\frac{\det K^{1/2}}{\pi^L})^{\frac{1}{4}} e^{-\frac{1}{2}<\phi|K^{1/2}|\phi>}.
\end{equation}
One can calculate the two point correlators $X_D=\Tr(\rho_D \phi_i \phi_j)$ and 
$P_D=\Tr(\rho_D \pi_i \pi_j)$ using the $K$ matrix, where $\rho_D$ is the reduced density matrix of domain $D$.
 The squared root of this matrix, as well as its inverse, can be split up  into 
coordinates of the subsystems $D$  and $\bar{D}$ , 
i. e., 
\begin{eqnarray}\label{X_D P_D}
 K^{-1/2}=
\begin{pmatrix}
  X_{D} & X_{D\bar{D}} \\
  X^{ T}_{D\bar{D}} & X_{\bar{D}} 
 \end{pmatrix}, \hspace{1cm} K^{1/2}=
\begin{pmatrix}
  P_{D} & P_{D\bar{D}} \\
  P^{ T}_{D\bar{D}} & P_{\bar{D}} 
 \end{pmatrix}.\nonumber
\end{eqnarray}
 The spectra of the  matrix $2C = \sqrt{X_DP_D}$, can be used to calculate              
 the  entanglement entropy, see  \cite{Casini2009a} and reference therein,

 \begin{equation}\label{EE harmonic oscillators}
S=\tr\Big{[}(C+\frac{1}{2})\log(C+\frac{1}{2})-(C-\frac{1}{2})\log(C-\frac{1}{2})\Big{]}.
\end{equation}
 Now if we do measurement on the position of all the oscillators $\{\phi_i\}\in A$ they will take some definite values (for example, $\{\phi_i\}$=0 ) and eventually  
will get decoupled from the rest
of the oscillators. In other words the final state will be the same as (\ref{GroundSwave}) but instead of $K^{1/2}$ we need to consider
$(K^{1/2})_{\bar{A}}$ which is a subblock of the matrix $K^{1/2}$ corresponding to the oscillators in the subsystem $\bar{A}$. This means that we now have a 
new Gaussian wave function and one can calculate its bipartite entanglement entropy
with the formula (\ref{EE harmonic oscillators}). The results for the short-range harmonic oscillators (discrete Klein-Gordon
field theory) are shown in the figure ~2 which show  perfect agreement with the equation (\ref{SB for PBC}) if we consider $c=1$ which is the central charge of
the free field theory. 
\begin{figure} [htb] \label{fig2}
\centering
\includegraphics[width=0.45\textwidth]{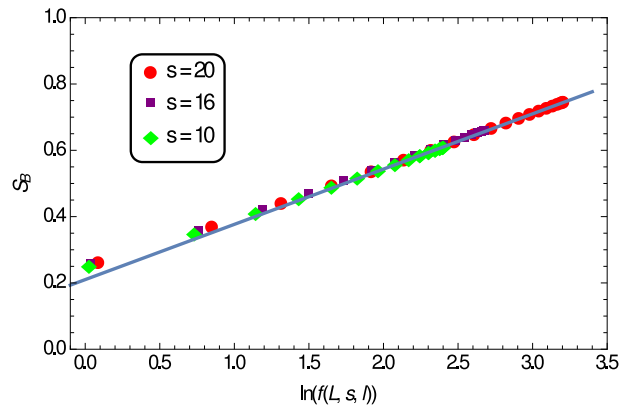}
\caption{Entanglement entropy of subregion $B$  for short-range coupled harmonic oscillators with total length $L=50$ and the measurement region sizes
$s=10,16$ and $20$ with respect to $\ln f (L,s,l)$, where 
$f (L,s,l)=\frac{L}{\pi}\frac{\sin\frac{\pi}{L}(l+s)\sin\frac{\pi}{L}l}{a\sin\frac{\pi}{L}s}$.
The full line is the function (\ref{SB for PBC}) with $c=1$ and  $\gamma_1=0.21$.} 
\end{figure}
Since the above results are independent of the outcome 
of the measurement of the the position of the oscillators $\{\phi_i\}\in A$
one can use them to find a lower bound for the localizable entanglement in this system as follows
\begin{eqnarray}\label{localizable}
S_{loc}(B,\bar{B}) > \frac{1}{6}\ln \Big{(}\frac{L}{\pi}\frac{\sin\frac{\pi}{L}(l+s)\sin\frac{\pi}{L}l}{a\sin\frac{\pi}{L}s}\Big{)}+\gamma_1.
\end{eqnarray}
Notice that the above result is  based on this fact that the scaling properties of the entanglement entropy after partial measurement in harmonic oscillators
is independent
of the outcome of the measurement. This is not necessarily correct in generic systems.

 \begin{figure} [htb] \label{fig3}
\centering
\includegraphics[width=0.45\textwidth]{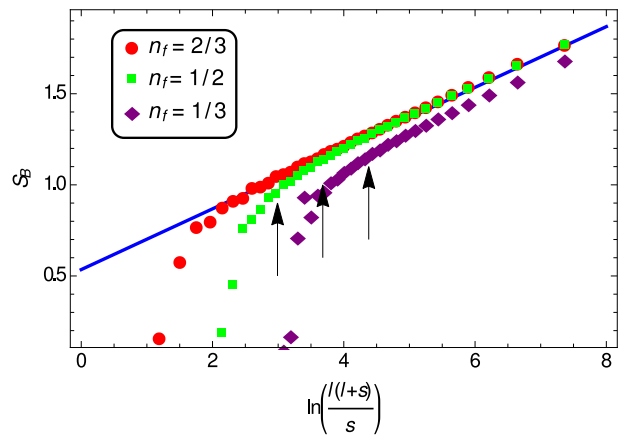}
\caption{Entanglement entropy of subregion $B$ with length $l$ after measurement on subsystem with length $s$ with ferromagnetic 
outcome for XX model in a magnetic field
. From top to bottom we have $n_f=\frac{2}{3},\frac{1}{2}$ and $\frac{1}{3}$. In our numerics $l+s=40$ is fixed.
The full line is the function (\ref{SB for PBC}) with $L\to \infty$ and $c=1$ and  $\gamma_1=0.53$ and the arrows are at $l^c_{n_f}=(1-n_f)(l+s)$
from left to right for
$n_f=\frac{2}{3},\frac{1}{2}$ and $\frac{1}{3}$ .} 
\end{figure}
\section{XX model in a magnetic field.}

In this section we would like to calculate bipartite entanglement entropy after partial projective measurement in XX model in a magnetic field. The method that we use 
can work equally for all the other spin chains that can be mapped to free fermions. The Hamiltonian of XX model is as follows:
\begin{equation}\label{XX model Hamiltonian}
H=-\frac{1}{2}\sum \Big{(}\sigma_l^x\sigma_{l+1}^x+\sigma_l^y\sigma_{l+1}^y-2h\sigma_l^z\Big{)}.
\end{equation}
After using simple Jordan-Wigner transformation, i.e. $c_l=\prod_{_{m<l}}\sigma_m^z\frac{\sigma_l^x+i\sigma_l^y}{2}$, the Hamiltonian will have the following form
\begin{equation}\label{XX model Hamiltonian free fermion}
H=-\sum \Big{(}c_l^{\dagger}c_{l+1}+c_{l+1}^{\dagger}c_l+2h(c_l^{\dagger}c_l-\frac{1}{2})\Big{)}.
\end{equation}
The entanglement entropy of a subsystem  can be calculated easily, see \cite{Peschela}, by first calculating the reduced density matrix 
$\rho_B=K e^{-\sum \tilde{H}_{ij}c_i^{\dagger}c_j}$ and then diagonalizing it, the final formula is
\begin{figure} [htb] \label{fig4}
\centering
\includegraphics[width=0.45\textwidth]{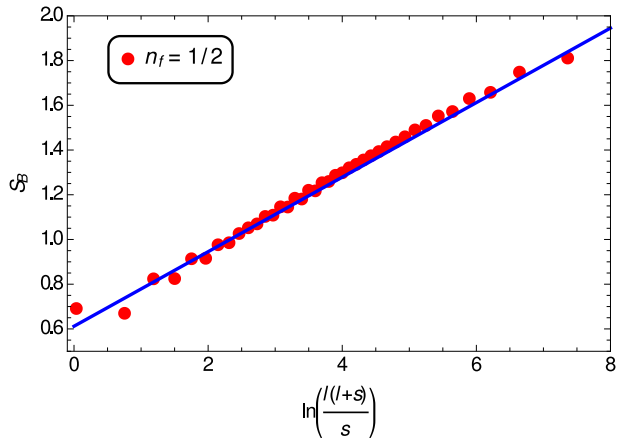}
\caption{Entanglement entropy of subregion $B$ with length $l$ after measurement on subsystem with length $s$ 
with antiferromagnetic outcome for XX model in half filling
. In our numerics $l+s=40$ is fixed.
The full line is the function (\ref{SB for PBC}) with $L\to \infty$ and $c=1$ and  $\gamma_1=0.61$.} 
\end{figure}
\begin{equation}\label{EE free fermions}
S=-\tr\Big{[}(1-C)\log(1-C)+C\log C\Big{]}.
\end{equation}
The matrix $\tilde{H}$ is related to the correlation matrix $C$ as
$\tilde{H}=\log(C^{-1}-1)$. In
our case the correlation matrix for infinite system has the form $C_{ij}=<c_i^{\dagger}c_j>=\frac{\sin(\pi n_f(i-j))}{\pi(i-j)}$, 
where $n_f=\frac{\arccos|h|}{\pi}$.
Looking to the Jordan-Wigner transformation it is easy to see that any measurement in the $\sigma^z$ basis on particular site can be translated to the measuring
the number of fermions in that site. In other words if one measures the $\sigma^z$ in all the sites of a subsystem with size $s$ the outcome
will be one of $2^s$ possible configurations which can be easily translated to the configurations made of presence or absence of fermions
in the sites of the subsystem. For simplicity we first consider that the outcome of the measurement is a string of $s$  fermion occupied sites.
This can be calculated easily by using Grassmann numbers by first calculating the reduced density matrix
for a sub-system with length $l+s$ and then finding the reduced density matrix of the subsystem with length $l$ with the assumption
that the outcome of the measurement in the subsystem $s$ is a string of filled sites. The method is explained with full detail in the appendix.
The results for different filling factors are shown in the Figure ~3. There are some comments in order: first of all because of the $U(1)$
symmetry of the XX model the number of particles in the system is conserved and for this reason as far as $n_f$ is small it is very difficult to have 
a string of sites with fermions. In other words for example in the XX model with $h=0$ the most probable outcome
is an antiferromagnetic string rather than ferromagnetic string. In \cite{Stephan2013} it is already conjectured that  for the half filling
case most probably the ferromagnetic configuration is not going to lead to a boundary conformal field theory and so the very first assumption that we 
used is going to fail. Despite this argument we found surprisingly that for  $l>(\frac{1}{n_f}-1)s$ the formula (\ref{SB for PBC}) 
works perfectly.

One can also check the results for those cases that the outcome of the measurement is antiferromagnetic string. It is expected that this case
leads to Dirichlet boundary condition in the bosonization language and so it is related the boundary CFT. The results presented in the Figure ~4
indeed show that the formula (\ref{SB for PBC}) works perfectly also in this case. Since the ferromagnetic and antiferromagnetic
configurations are at the two extreme sides of the all possible configurations based on the above numerical results
one can conjecture that independent of the outcome of the measurement in the $\sigma^z$ basis the formula (\ref{SB for PBC}) will work if the $l$ is 
 bigger enough than $s$.
Note that Since the number of configurations increases exponentially this conjecture is  beyond what we can check numerically.

\section{Open boundary conditions} 

One can also do all the CFT calculations  in the case of open boundary condition, the only difference is that now we have a strip with a slit instead of a cylinder.
One can simply map a strip with a slit to upper half plane by using the map $z(w)=\sqrt{1-\frac{\tan^2\frac{\pi w}{2L}}{\tan^2\frac{\pi s}{2L}}}$. After
some algebra we have
\begin{eqnarray}\label{SB for OBC}
S_B=\frac{c}{6}\ln \Big{(}\frac{2L}{\pi}\frac{\cos\frac{\pi s}{L}-\cos\pi\frac{l+s}{L}}{a\cos^2\frac{\pi s}{2L}}\cot\frac{\pi(l+s)}{2L}\Big{)}+\gamma_{2}.
\end{eqnarray}
The limit $s\to 0$ simply gives us the well-known formula $S_B=\frac{c}{6}\ln(\frac{L}{\pi}\sin\frac{\pi l}{L})$, see \cite{Calabrese2004}. 
We have checked the validity of this formula in the case of short-range harmonic oscillators using the same method that we hired 
for the periodic BC. The results shown in Figure ~5 are fairly compatible with the formula (\ref{SB for OBC}). We noticed that
although the log formula is perfectly compatible there is almost 8 percent deviation from central charge that might be due to finite size effects. 
\begin{figure} [htb] \label{fig5}
\centering
\includegraphics[width=0.45\textwidth]{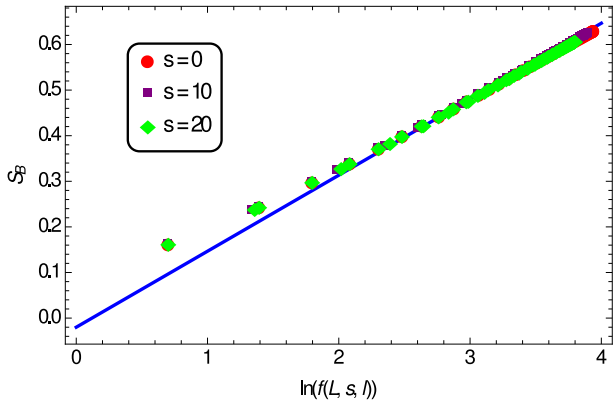}
\caption{Entanglement entropy of subregion $B$  for a system (short-range coupled harmonic oscillators)  with total length $L=50$ and the measurement region sizes
$s=0,10$ and $20$ with respect to $\ln f (L,s,l)$, where 
$f (L,s,l)=\frac{2L}{\pi}\frac{\cos\frac{\pi s}{L}-\cos\pi\frac{l+s}{L}}{a\cos^2\frac{\pi s}{2L}}\cot\frac{\pi(l+s)}{2L}$. 
The full line is the function (\ref{SB for OBC}) with $c=1$ and  $\gamma_2=-0.02$.} 
\end{figure}

\section{Conclusions.}
We described a general set up for calculating bipartite entanglement entropy after local projective measurement in
critical one dimensional quantum systems. Exact formulas were derived for bipartite entanglement entropy after ``conformal measurements``
in the case of periodic and open boundary conditions. The formulas were checked in explicit examples of 
free bosonic system and XX model in a magnetic field. We noticed that since bosonization of
XXZ model in the $\sigma^z$ basis leads to free  bosonic system the antiferromagnetic outcome of the measurement
in this basis should be compatible with the results of free bosonic system \cite{Stephan2013}. In the case
of XX model in a magnetic field we showed that if we do our measurements in the $\sigma^z$ basis independent
of having ferromagnetic or antiferromagnetic outcome for our measurements the bipartite entanglement entropy
in particular regimes can be described with CFT formulas. We also derived a lower bound for the localizable
entanglement in the case of harmonic oscillators. There are interesting questions remained to be answered: first of all
in our free fermion approach we were able to handle just $\sigma^z$ basis, it is important to look to the other
basis especially $\sigma^x$ basis by using exact diagonalization methods. Another interesting question is
investigating the same questions in the case of non-critical quantum systems and the validity of the area law. 
Some of these questions will be discussed in a forthcoming paper \cite{Rajabpour}. Finally understanding
the problem in the holographic set up \cite{Ryu} will surely help to extend some of these results to higher dimensions.

\paragraph*{Acknowledgments.} We  thank M. G. Nezhadhaghighi for early collaboration on the subject.

\setcounter{equation}{0}
\renewcommand{\theequation}{S\arabic{equation}}

\twocolumngrid

\begin{center}
{\Large Appendix}
\end{center}

\section*{Bipartite entanglement entropy after partial measurement of fermion occupation numbers in free fermions}

In this supplementary note we would like to present a method which  can be used to measure the entanglement entropy 
after measuring the fermion occupation numbers in a subsystem in a generic free fermion system. The method is based on Grassmann numbers
and it is a generalization of the work \cite{Peschela}. Although the method can be also used  for the most generic free fermion system \cite{Rajabpour}
here we will concentrate on a system with the following reduced density matrix:
\begin{eqnarray}\label{reduced density matrix}
\rho_B=K e^{-\sum \tilde{H}_{ij}c_i^{\dagger}c_j}
\end{eqnarray}
The matrix $\tilde{H}$ is related to the correlation matrix $C$ as
$\tilde{H}=\log(C^{-1}-1)$. Finally one can find the entanglement entropy using the following formula, see for example \cite{Casini2009a};
\begin{eqnarray}\label{entanglement entropy2}
S_B=-\tr\Big{[}(1-C)\log(1-C)+C\log C\Big{]}.
\end{eqnarray}
To calculate the reduced density matrix after measurement we need to first define fermionic coherent states. It can be defined as follows

\begin{eqnarray}\label{fermionic coherent states1}
 |\boldsymbol{\xi}>= |\xi_1\xi_2...\xi_N>= e^{-\sum_{i=1}^N\xi_ic_i^{\dagger}}|0>,
\end{eqnarray}
where $\xi_i$'s are Grassmann numbers following the  properties: $\xi_n\xi_m+\xi_m\xi_n=0$ and $\xi_n^2=\xi_m^2=0$. Then it is easy to show that
\begin{eqnarray}\label{fermionic coherent states2}
c_i |\boldsymbol{\xi}>= -\xi_i  |\boldsymbol{\xi}>.
\end{eqnarray}
Using the above formula one can derive the following useful formula:
\begin{eqnarray}\label{matrix elements1}
<\boldsymbol{\xi}|  e^{\sum_{ij}A_{ij}c_i^{\dagger}c_j} |\boldsymbol{\xi'}>=e^{\sum_{ij}(e^A)_{ij}\xi_i^*\xi_j'}.
\end{eqnarray}
With the same method one can also define another kind of fermionic coherent state as 
\begin{eqnarray}\label{fermionic coherent states3}
 |\boldsymbol{\eta}>= |\eta_1\eta_2...\eta_N>= e^{-\sum_{i=1}^N\eta_ic_i}|1>,
\end{eqnarray}
where $\eta_i$'s are Grassmann numbers. Then it is easy to show that
\begin{eqnarray}\label{fermionic coherent states4}
c_i^{\dagger} |\boldsymbol{\eta}>= -\eta_i  |\boldsymbol{\eta}>
\end{eqnarray}
and consequently
\begin{eqnarray}\label{matrix elements2}
<\boldsymbol{\eta}|  e^{\sum_{ij}A_{ij}c_i^{\dagger}c_j} |\boldsymbol{\eta'}>=e^{\sum_{ij}(e^{-A})_{ij}\eta_i^*\eta_j'}.
\end{eqnarray}

The two formulas (\ref{matrix elements1}) and (\ref{matrix elements2}) are the main formulas one can use to calculate reduced density matrix
after measurement. We first discuss how the method works for a string of empty sites. A string of empty sites means that in the equation (\ref{matrix elements1})
we need to put all the $\xi_i$'s with $i$ inside the measuring subsystem equal to zero. This is equivalent  to keeping just all 
the elements of the matrix $e^A$ outside the subsystem. Then we take the logarithm of this matrix and find a new matrix $\tilde{A}$ which is going to be
our new entanglement Hamiltonian and then one can easily calculate the entanglement entropy of that system
with the method we explained at the beginning of this section. To do the same calculation for an string of occupied sites
instead of equation  (\ref{matrix elements1})    one needs to use equation (\ref{matrix elements2}) and follow the same method as we just explained.
In principle this method can be applied for any possible configuration by just appropriate using the equations 
(\ref{matrix elements1}) and (\ref{matrix elements2}).




\end{document}